\documentstyle[multicol,epsf,aps]{revtex}
\begin{document}
\preprint{SNUTP 99-023}
\title{Lattice effects on the current-voltage
characteristics of superconducting arrays}
\author{Mina Yoon and M.Y. Choi}
\address{Department of Physics and Center for Theoretical Physics, 
 Seoul National University, Seoul 151-742, Korea}
\author{Beom Jun Kim}
\address{Department of Theoretical Physics, Ume{\aa} University, S-901 87 Ume{\aa}, Sweden}
\maketitle
\draft
\begin{abstract}
The lattice effects on the current-voltage characteristics of two-dimensional arrays of 
resistively shunted Josephson junctions are investigated.
The lattice potential energies due to the discrete lattice structure
are calculated for several geometries and directions of current injection.
We compare the energy barrier for vortex-pair unbinding 
with the lattice pinning potential, which shows that
lattice effects are negligible in the low-current limit as well as in the 
high-current limit. 
At intermediate currents, on the other hand, 
the lattice potential becomes comparable to the barrier height
and the lattice effects may be observed in
the current-voltage characteristics.
\end{abstract}

\pacs{PACS number: 74.50.+r, 74.25.Fy, 74.60.Ge}

\begin{multicols}{2}

Two-dimensional (2D) arrays of weakly coupled Josephson junctions in equilibrium
are well described by the 2D $XY$ model, which exhibits the Berezinskii-Kosterlitz-Thouless (BKT) 
transition driven by the unbinding of vortex-antivortex pairs~\cite{intro}. 
In experiment, on the other hand, the systems are usually in nonequilibrium and  
dynamical quantities are measured in the presence of external driving~\cite{intro,intro1}. 
Although the static nature of the BKT transition has been well established, 
there still remain unsolved questions on the dynamics of the system, 
for example, the value of the exponent of the current-voltage ($IV$) characteristics, 
the dynamic universality class, and the noise spectrum~\cite{simkin,Kim,tiesinga}. 
These 2D Josephson junction arrays (JJAs) also draw much interest in relation with 
superconducting films and highly anisotropic high-$T_c$ superconductors. 
In JJAs, unlike the latter superconducting materials, the underlying discrete
lattice structure causes nonzero pinning potentials for vortices;
such lattice pinning potentials have not been taken into account properly in the theoretical 
studies based on the continuum limit.

This work investigates the lattice effects in the 2D $N \times N$ resistively-shunted 
junction (RSJ) model. We first calculate the lattice potential in several geometries 
including different lattice structures and directions of current injection. 
The lattice potential barrier on a square array with diagonal current injection is found 
to be much larger than that on the same array with horizontal current injection.  
Comparing the obtained lattice pinning potential with the energy barrier 
which a bound vortex pair should overcome to be free vortices, 
we find that the effects of lattice pinning are negligibly small in the low-current
regime as well as in the high-current regime;
this is confirmed by the $IV$ characteristics computed numerically 
for the square arrays with different directions of current injection.
It is also found that there exists an intermediate regime of current, where
the lattice potential effects are observable.

\begin{figure}
\centerline{\epsfxsize=8cm \epsfbox{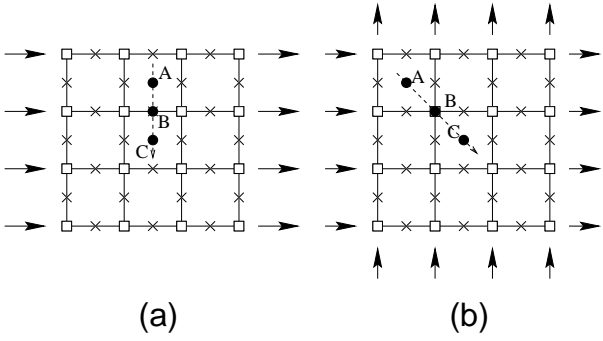}}
\vskip 0.5cm
\caption {Illustration of the positions of a vortex under currents injected 
 (a) in the $x$ direction and (b) in both $x$ and $y$ directions.
 The positions $A$ and $C$ correspond to the local minimum energy $E_{\rm min}$,
 while $B$ corresponds to the maximum $E_{\rm max}$. 
 The dashed lines with arrows describe paths of the vortex under the currents.
 Squares represent superconducting islands and arrows denote current injection.}
\label{fig:jja}
\end{figure}

\begin{figure}
\centerline{\epsfxsize=8cm  \epsfbox{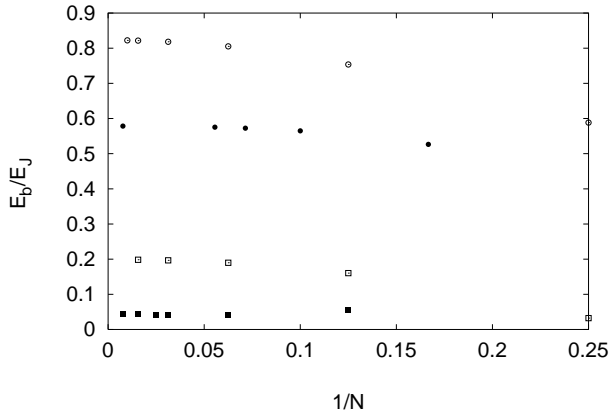}}
\vskip 0.5cm
\caption {The lattice potential barrier $E_b$ versus the inverse system size $1/N$.
Filled squares, open squares, and filled circles correspond to the 
triangular, square, and honeycomb arrays with horizontal current injection, respectively; 
open circles to the square array with diagonal current injection.}
\label{fig:barrier}
\end{figure}

We begin with the calculation of the lattice potential energy, which a vortex should 
overcome to move to the next face of the lattice.
In the presence of external currents, a vortex is expected to be exerted by the 
Magnus force 
\begin{equation} \label{eq:magnus}
{\bf F} = \frac{1}{c} \Phi_0 {\bf J} \times \hat{\bf z},
\end{equation}
where ${\bf J}$ is the external current density and $\Phi_0 \equiv h c /2e$ 
is the flux quantum.
Here the direction of the Magnus force is perpendicular to that of the current density,
enforcing the vortex to move along the perpendicular direction 
to the current injection.
As an example, Fig.~\ref{fig:jja} displays square arrays in the presence of 
the external currents (a) in the horizontal and (b) in the diagonal direction. 
Under the (sufficiently strong) Magnus force, the vortex at position A,
which corresponds to the phase configuration with local minimum energy $E_{\rm min}$, 
passes position $B$, with maximum energy $E_{\rm max}$, and
then moves to $C$, which yields the same configuration as $A$.
Accordingly, the potential barrier $E_b$ that the vortex should overcome to move
is determined by~\cite{Lobb} 
\begin{equation}
E_b = E_{\rm max} - E_{\rm min}
\label{eq:BH}
\end{equation}
with the energy 
\begin{equation}
E = -E_J \sum_{\langle i j\rangle} \cos(\phi_i - \phi_j),
\label{eq:ES}
\end{equation}
where $E_J$ is the Josephson coupling strength, $\phi_i - \phi_j$ is the phase 
difference between sites $i$ and $j$, and the summation is taken over all
nearest neighboring pairs.  
We use the method in Ref.~\onlinecite{Lobb} to obtain the phase configuration:
The minimization of the energy in Eq.~(\ref{eq:ES})
leads to the condition that the net current flowing into site $i$ should vanish:
\begin{equation}
\sum_j \sin( \phi_i - \phi_j ) = 0, 
\label{eq:zeroI1}
\end{equation}
which can be rewritten as
\begin{equation}
\tan\phi_i = \frac{ \sum_j \sin \phi_j }{ \sum_j \cos \phi_j}
\label{eq:zeroI2}
\end{equation}
with the summations performed over the four nearest neighbors of $i$. 
The phase configuration can then be found iteratively  from
\begin{equation}
\tan\phi_i^{(n+1)} = \frac{\sum_j \sin \phi_j^{(n)}}{ \sum_j \cos \phi_j^{(n)}},  
\label{eq:iteration}
\end{equation}
where $\phi_j^{(n)}$ is the value of $\phi_j$ at the $n$th iteration.
Equation~(\ref{eq:iteration}), together with appropriate boundary conditions, 
gives the phase configurations for the minimum energy 
$E_{\rm min}$ and for the maximum energy $E_{\rm max}$. 
In this manner we compute the lattice potential barrier for $N \times N$
square, triangular, and honeycomb arrays with horizontal current injection, 
as well as for a square array with diagonal current injection.
Figure~\ref{fig:barrier} shows the obtained lattice potential barrier $E_b$ versus 
the inverse size $1/N$, which manifests that the barrier heights saturate to constant values
for sufficiently large $N$.
In the case of square and  triangular arrays with horizontal current injection,
the barrier heights saturate to the values $E_b/E_J \approx 0.199$ and $0.043$, 
reproducing the results obtained in Ref.~\onlinecite{Lobb}. 
In addition, Fig.~\ref{fig:barrier} also
gives the values $E_b/E_J \approx 0.575$ and $0.822$ for a honeycomb array
with horizontal current injection and for a square array with diagonal 
injection, respectively.

\begin{figure}
\centerline{\epsfxsize=8cm  \epsfbox{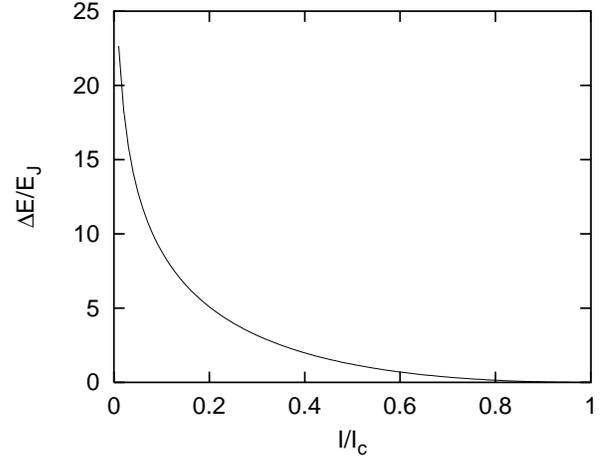}}
\vskip 0.5cm
\caption{The energy barrier $\Delta E$ for pair unbinding
versus external current $I$ in the absence of lattice pinning effects. 
It is shown that $\Delta E = 0$ for $I \geq I_c$.} 
\label{fig:DE} 
\end{figure}

We first consider the energy for unbinding of a vortex pair 
without the lattice pinning potential and then examine 
how the result changes as the pinning effects are included.
The interaction energy of a vortex-antivortex pair 
separated by distance $r$ is given by $E_1 G'(r)$,
where $E_1 \approx 2\pi E_J$ at low temperatures and 
$G'(r)$ is the lattice Coulomb Green function (with the diagonal part subtracted)~\cite{KT}.
To an excellent approximation, $G'(r)$ takes the form
\begin{equation}
G'(r) \approx \ln (r/a) + C
\end{equation}
for all $r\geq a$, 
where $a$ is the lattice spacing and $C$ is a constant~\cite{Spitzer}.
Accordingly, in the presence of external current $I$ 
the energy of a vortex-antivortex pair reads~\cite{Weber}
\begin{equation} 
E(r) = E_0 + E_1 \ln \left(\frac{r}{a}\right) - \left(\frac{hI}{2ea}\right) r ,
\label{eq:vortexE}
\end{equation}
where $E_0$ is a constant
and the last term arises from the Magnus force in Eq.~(\ref{eq:magnus}). 
We measure the energy in units of $E_J$ and 
write Eq.~(\ref{eq:vortexE}) in the dimensionless form:
\begin{equation} 
E(r) =  E_1 \ln r - 2\pi I r,
\label{eq:vortexE1}
\end{equation}
where $E_0$ in Eq.~(\ref{eq:vortexE1}) has been dropped for convenience,
and $r$ and $I$ are in units of $a$ and the single junction
critical current $I_c \equiv 2eE_J/\hbar$, respectively.
The condition for the maximum pair energy $\partial E/\partial r = 0$
gives the estimation of the maximum pair size:
\begin{equation}
r_{\rm max} = \frac{E_1}{2\pi I}
\end{equation} 
for $I \leq E_1/2\pi \approx 1$ (in the dimensionless form).  
(For $I \gtrsim 1$, we have $r_{\rm max}\approx 1$.)
The energy barrier $\Delta E$ for the pair unbinding is thus given by
\begin{equation}
\Delta E = E(r_{\rm max}) - E(1) = E_1\left[\ln\left( \frac{E_1}{2\pi I}\right) - 1\right] + 2\pi I
\label{eq:vortexmaxE}
\end{equation}
in the absence of the lattice pinning potential.
We display $\Delta E$ as a function of $I$ in Fig.~\ref{fig:DE}, where it
is observed that $\Delta E$ has very large values in the small-current regime. 
Note also that $\Delta E = 0$ for $I \geq 1$, which implies that 
the vortex-antivortex pair can unbind even at zero temperature 
if the external current is larger than the critical current. 

We now consider the lattice pinning effects on the vortex-antivortex pair energy. 
Taking the position of the vortex as the origin, we write
the lattice pinning potential in the simple form
\begin{equation}
E_{p}(r) = -\frac{E_b}{2} \cos 2\pi r,
\end{equation}
which is to be included in Eq.~(\ref{eq:vortexE1}), 
with the lattice potential barrier $E_b$ in units of $E_J$. 
In the presence of such a lattice potential, 
the vortex may feel some roughness when it moves around.  
Under small or large external currents, however, 
the lattice effects on the vortex motion are expected not to be appreciable:
In the latter case of large currents,
the lattice pinning potential $E_p$ is so small compared with
the driving potential $-2\pi Ir$ in Eq.~(\ref{eq:vortexE1}),
thus resulting in negligible effects.
At small external currents, the energy barrier $\Delta E$ for the pair unbinding,
shown in Fig.~\ref{fig:DE},
is much larger than the lattice potential barrier $E_b$,
and dominates the transport properties of vortices 
since the vortex-antivortex pair should overcome the largest 
energy barrier to be free vortices. 
On the other hand, it is of interest to note that there exists the intermediate-current regime,
where $E_b$ is comparable to $\Delta E$.
In that regime
the lattice effects on transport properties such as the $IV$ characteristics
can presumably be measured.

To investigate the above possibility, we focus on square arrays in the presence
of external currents in horizontal and diagonal directions and compute
the $IV$ characteristics at finite temperatures. Since these two cases
(see Fig.~\ref{fig:jja}) are believed to differ only in the lattice potential
barrier, we expect that any difference in the $IV$ characteristics should be 
attributed to the lattice pinning effects.
The net current through a Josephson junction of shunt resistance $R$
is given by the sum of the supercurrent, normal
current, and thermal noise current;
the resulting current conservation condition at each grain
yields the equations of motion for a square $N\times N$ JJA~\cite{intro,simkin}:
\begin{equation}
\label{eq:dyn1}
I_i^{\rm ext} = {\sum_j}' \left[ {\hbar \over 2eR} {d \over dt}
(\phi_i - \phi_j ) + I_c \sin(\phi_i - \phi_j ) + \eta_{ij}\right],
\end{equation}
where $I_i^{\rm ext}$ is the external current fed into grain $i$,
the primed summation runs over the nearest neighbors of grain $i$,
and $\eta_{ij}$ is the thermal noise current.
We here employ the fluctuating boundary conditions, and
introduce the twist variables ${\bf \Delta} \equiv
(\Delta_x,\Delta_y)$ to write the phase difference
between the nearest-neighboring grains in the form
\begin{displaymath}
\phi_i - \phi_j - {\bf r}_{ij} \cdot {\bf \Delta},
\end{displaymath}
where ${\bf r}_{ij}$ is the displacement between $i$ and $j$
and the periodic boundary conditions on $\{\phi_i\}$ are imposed in both
directions.
The equations of motion in Eq.~(\ref{eq:dyn1}) then take the form~\cite{Kim}
\begin{equation}\label{eom}
{\sum_j}' \left[ {d \over dt} (\phi_i - \phi_j )   \label{eq:dyn2}
 + \sin(\phi_i - \phi_j - {\bf r}_{ij} \cdot {\bf \Delta}) + \eta_{ij}\right]=0,
\end{equation}
where
time has been rescaled in units of $\hbar/2eRI_c$, and
the thermal noise current in units of $I_c$ satisfies 
$\langle \eta_{ij}(t) \rangle = 0$ and $\langle \eta_{ij}(t) \eta_{kl} (0)\rangle
= 2T(\delta_{ik}\delta_{jl} -  \delta_{il}\delta_{jk})\delta(t)$ with 
temperature $T$ in units of $E_J/k_B$.
The dynamics of the twist variables is governed by
two additional equations: 
\begin{eqnarray}
\dot\Delta_x &=& \frac{1}{N^2} \sum_{\langle ij\rangle_x}
    \sin(\phi_i - \phi_j - \Delta_x) +\eta_{\Delta_x} - I_x  
    \nonumber \\
\dot\Delta_y &=& \frac{1}{N^2} \sum_{\langle ij\rangle_y}
     \sin(\phi_i - \phi_j - \Delta_y) +\eta_{\Delta_y} -I_y   \label{eq_Deltay_Id}, 
\end{eqnarray}
where the summation $\sum_{\langle ij \rangle_x}$ is over all links in the $x$ direction
and thermal noise terms satisfy 
$\langle \eta_{\Delta}(t) \rangle = 0$ and $\langle \eta_{\Delta}(t) 
\eta_{\Delta}(0) \rangle = (2T/N^2)\delta(t)$. 
In the case of horizontal current injection, we have
$I_x = I$ and $I_y = 0$, while for diagonal injection $I_x = I_y = I/\sqrt{2}$.
By means of the Euler algorithm, we integrate Eq.~(\ref{eq_Deltay_Id}) 
and compute the voltage $V = \sqrt{V_x^2 + V_y^2}$, where
$V_{x(y)} \equiv -N \langle \dot\Delta_{x(y)} \rangle_t$ 
with $\langle \cdots \rangle_t$ denoting the time average. 
The use of the fluctuating twist boundary conditions
in the presence of external currents has the advantage that
the direction of current injection can be controlled easily.

\begin{figure}
\centerline{\epsfxsize=8cm  \epsfbox{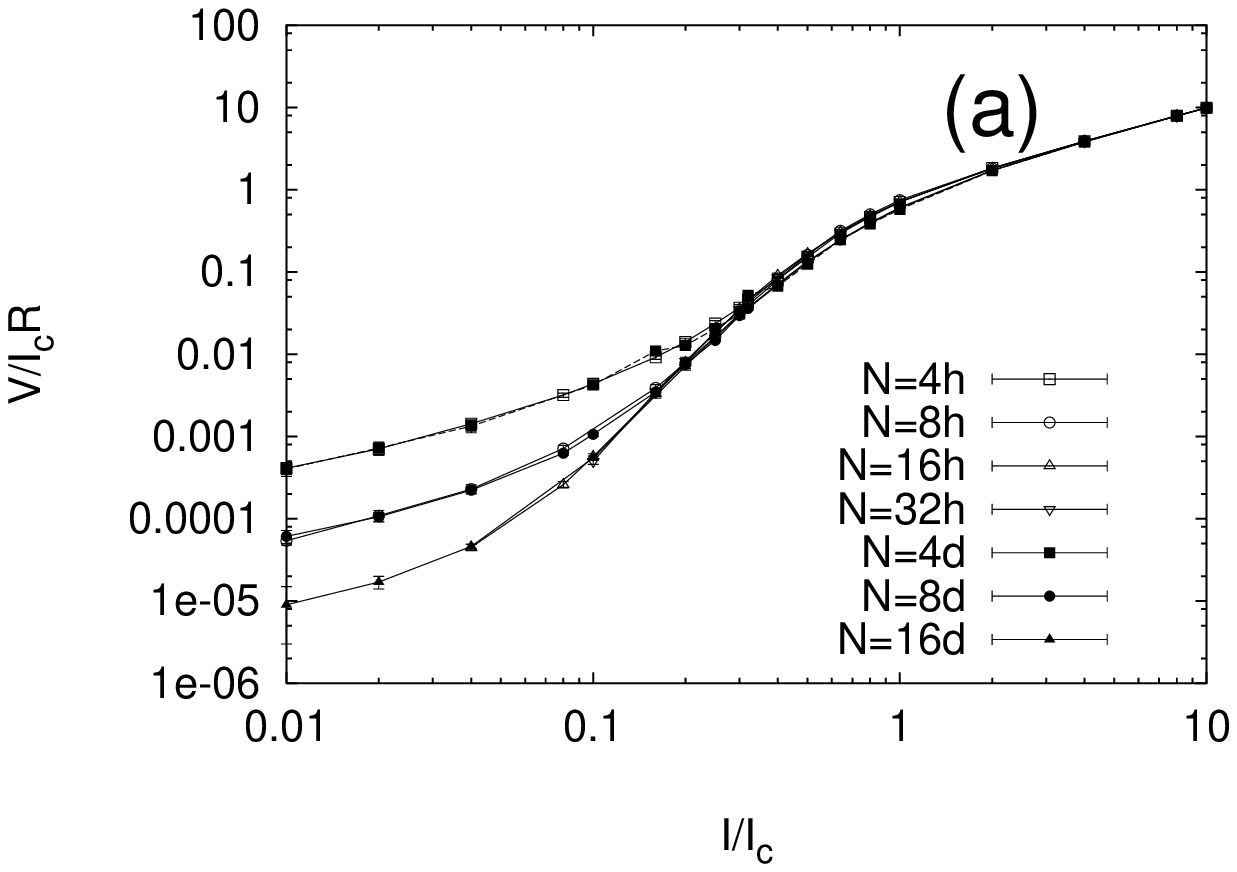}}
\centerline{\epsfxsize=8cm  \epsfbox{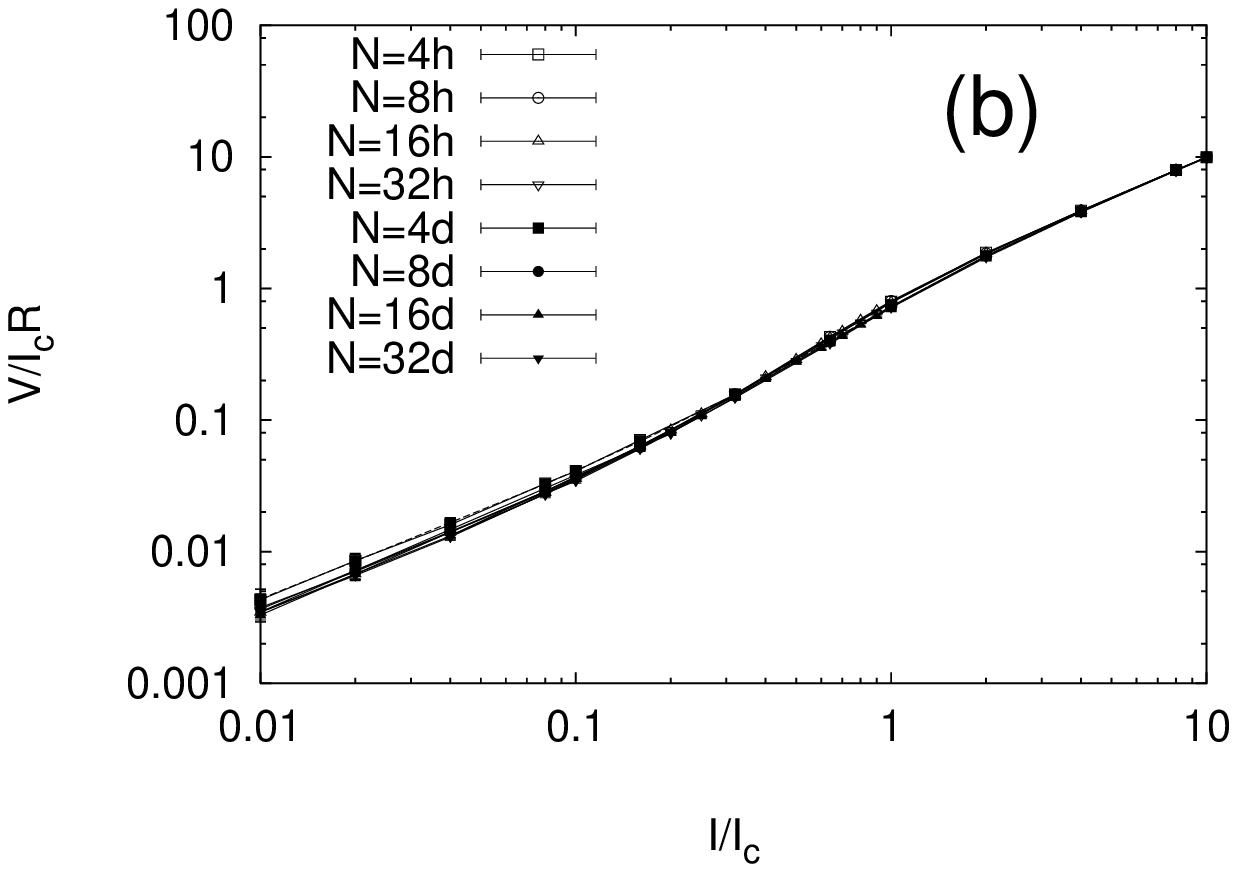}}
\vskip 0.5cm
\caption
{Current-voltage characteristics of square arrays of sizes $N=4, 8, 16$ and 32
 at (a) $T=0.84$ and (b) at $T=1.30$.
 For comparison, the data for horizontal current injection (labeled by $h$) 
 and those for diagonal current injection (labeled by $d$) are plotted together.}
\label{fig:iv}
\end{figure}

\begin{figure}
\centerline{\epsfxsize=8cm  \epsfbox{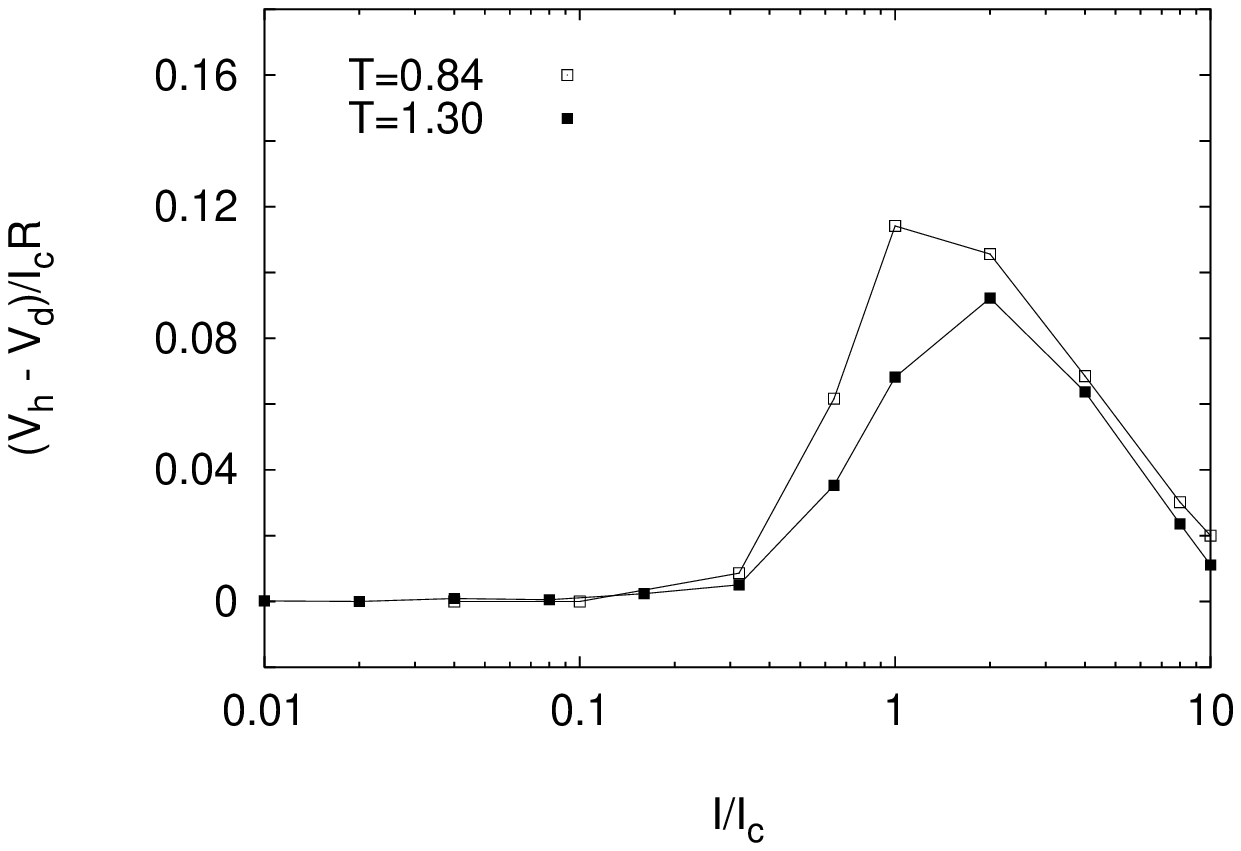}}
\vskip 0.5cm
\caption[Difference in the voltage between the two types of current injection
at $T=0.84$ and $1.30$]
{Difference between the voltage under
horizontal injection ($V_h$) and that under diagonal injection ($V_d$)
versus current in the system of size $N=16$ at $T=0.84$ and 1.30.
The lattice potential effects on the current-voltage characteristics are
shown to be appreciable only in the intermediate-current regime.}
\label{fig:diff}
\end{figure}

Figure~\ref{fig:iv} presents the resulting $IV$ curves of the arrays of sizes
$N = 4,8,16$, and $32$ under horizontal and diagonal current injections
at (a) $T=0.84$ and (b) $T=1.30$. As expected from the existence
of the resistive BKT transition at $T=T_{\rm BKT} \approx 0.9$, 
the voltage $V$ at $T=0.84$ (below $T_{\rm BKT}$) 
keeps decreasing as the system size is increased in the low-current regime, 
while $V$ appears to saturate to a nonzero value at $T=1.30$.
It is, however, rather difficult to discern the data for horizontal and diagonal
injection in the logarithmic scale.
To manifest the difference and to reveal the lattice effects in detail, 
we thus plot in the linear scale
the difference between the voltage under
horizontal injection ($V_h$) and that under diagonal injection ($V_d$),
which is displayed in Fig.~\ref{fig:diff}
as a function of the current $I$ in the system of size $N=16$ at $T=0.84$ and $1.30$. 
It is obvious that the difference $V_h{-}V_d$ indeed approaches zero in the limit of small 
and large currents, confirming the previous prediction based on the
comparison of the energy scales: the lattice pinning potential and the energy
barrier for pair unbinding. 
In particular, independence of the voltage in the low-current regime upon
the direction of current injection suggests that
the $IV$ exponent for a square array with horizontal
current injection found in Ref.~\onlinecite{Kim} is universal 
in the sense that it is independent of the underlying lattice structure. 
Furthermore, since the lattice pinning potential
is much smaller for horizontal injection ($E_b \approx 0.199 E_J$) than for 
diagonal injection ($E_b \approx 0.822 E_J$), vortices with horizontal injection
can move around more freely, which implies that in the intermediate-current
regime we have $V_h > V_d$, as again confirmed in Fig.~\ref{fig:diff}.

In conclusion, we have studied the lattice effects in two-dimensional 
arrays of resistively shunted Josephson junctions for several geometries
and directions of current injection. The lattice potential energy
due to the underlying discrete array structure has been calculated and 
then compared with the energy scale which a vortex-antivortex pair should 
overcome to be unbound. From this comparison, we have found that 
lattice pinning effects can be observed in the intermediate-current 
regime, which has been confirmed by the direct numerical
integration of the current-voltage characteristics.

We thank G.S. Jeon and P. Minnhagen for useful discussions, and
acknowledge the partial support from the SNU research fund,
from the Korea Research Foundation, and from the Korea Science and Engineering
Foundation.

\narrowtext

\newpage

\end{multicols}


\begin{references}

\bibitem{intro}
For a list of references, see, e.g., 
{\it KT Transitions and Superconducting Arrays}, edited by D. Kim, J.S. Chung, and
M.Y. Choi (Min Eum Sa, Seoul 1993);
{\it Macroscopic Quantum Phenomena and Coherence in Superconducting Networks},
edited by C. Giovannella and M. Tinkham (World Scientific, Singapore, 1995);
Physica B {\bf 222}, 253 (1996).

\bibitem{intro1}
B.J. van Wees, H.S.J. van der Zant, and J.E. Mooij, Phys. Rev. B {\bf 35}, 7291 (1987);
J.D. Carini, {\it ibid.} {\bf 38}, 63 (1988);
H.S.J. van der Zant, H.A. Rijiken, and J.E. Mooij, J. Low. Temp. Phys. {\bf 79}, 289 (1990).

\bibitem{simkin} 
M.Y. Choi and S. Kim, Phys. Rev. B {\bf 44}, 10411 (1991);
S. Kim and M.Y. Choi, {\it ibid.} {\bf 48}, 322 (1993);
M.V. Simkin and J.M. Kosterlitz, {\it ibid}. {\bf 55}, 11646 (1997).

\bibitem{Kim} B.J. Kim, P. Minnhagen, and P. Olsson, Phys. Rev. B {\bf 59}, 11506 (1999).

\bibitem{tiesinga} P.H.E. Tiesinga, T.J. Hagenaars, J.E. van Himbergen, and J.V. Jos\'{e}, 
Phys. Rev. Lett. {\bf 78}, 519 (1997); I.-J. Hwang and D. Stroud, Phys. Rev. B {\bf 57}, 
6036 (1998).

\bibitem{Lobb}
C.J. Lobb, D.W. Abraham, and M. Tinkham, Phys. Rev. B {\bf 27}, 150 (1983).

\bibitem{KT}
J.M. Kosterlitz and D.J. Thouless, J. Phys. C {\bf 6}, 1181 (1973);
J.V. Jos\'e, L.P. Kadanoff, S. Kirkpatrick, and D.R. Nelson, Phys. Rev. B {\bf 16}, 1217 (1977).

\bibitem{Spitzer}
See, e.g., F. Spitzer, {\it Principles of Random Walk} (Van Nostrand, Princeton, 1964),
pp.~148-151.

\bibitem{Weber}
V. Ambegaokar, B.I. Halperin, D.R. Nelson, and E.D. Siggia, Phys. Rev. B {\bf 21}, 1806 (1980);
H. Weber, M. Wallin, and H.J. Jensen, {\it ibid}. {\bf 53}, 8566 (1996).


\end{references}
\end{document}